\documentclass[reprint,superscriptaddress,nofootinbib,twocolumn,amsmath,amssymb,aps,prl]{revtex4-2}

\usepackage{graphicx}
\usepackage{dcolumn}
\usepackage{float}
\usepackage{bm}
\usepackage[colorlinks]{hyperref}
\usepackage[utf8]{inputenc}	
\usepackage{lmodern} 
\usepackage{siunitx}
\usepackage[version=4]{mhchem}
\usepackage{microtype}

\newcommand{\braket}[1]{\langle#1\rangle}

\begin{document}
\title{The Quantum Gaussian-Schell Model: A Link Between Classical and Quantum Optics}

\author{Riley B. Dawkins}
\affiliation{Quantum Photonics Laboratory, Department of Physics \& Astronomy, Louisiana State University, Baton Rouge, LA 70803, USA}

\author{Mingyuan Hong}
\affiliation{Quantum Photonics Laboratory, Department of Physics \& Astronomy, Louisiana State University, Baton Rouge, LA 70803, USA}

\author{Chenglong You}
\email{cyou2@lsu.edu}
\affiliation{Quantum Photonics Laboratory, Department of Physics \& Astronomy, Louisiana State University, Baton Rouge, LA 70803, USA}

\author{Omar S. Maga\~na-Loaiza}
\affiliation{Quantum Photonics Laboratory, Department of Physics \& Astronomy, Louisiana State University, Baton Rouge, LA 70803, USA}

\date{\today}

\begin{abstract}

The quantum theory of the electromagnetic field uncovered that classical forms of light were indeed produced by distinct superpositions of nonclassical multiphoton wavepackets. Specifically, partially coherent light represents the most common kind of classical light. Here, for the first time, we demonstrate the extraction of the constituent multiphoton quantum systems of a partially coherent light field. We shift from the realm of classical optics to the domain of quantum optics via a quantum representation of partially coherent light using its complex-Gaussian statistical properties. Our formulation of the quantum Gaussian-Schell model unveils the possibility of performing photon-number-resolving detection to isolate the constituent quantum multiphoton wavepackets of a classical light field. We experimentally verified the coherence properties of isolated vacuum systems and wavepackets with up to sixteen photons. Our findings not only demonstrate the possibility of observing quantum properties of classical macroscopic objects, but also establish a fundamental bridge between the classical and quantum worlds.

\end{abstract}

\maketitle


The work performed by Allan Schell in 1961 shaped the field of optical physics and specifically the classical theory of optical coherence \cite{schell1961multiple,Born1999}. His approach to describe spatial coherence of classical light fields laid the foundations for the development of optical technology ranging from imaging instruments and spectroscopy to communication \cite{Zhang2021, Dutta2014, Wheeler2011, chen2021quantum, Liu2022a}. This model, now known as the Gaussian-Schell model (GSM), enables describing classical coherence of optical wavefronts with different polarization and spectral properties \cite{wolf2007introduction, chen2021quantum, Peng2018, Yan2022, Liu2022a, Dutta2014, Rodenburg2014}. Interestingly, this model also enables modeling of propagating light in complex media \cite{Agrawal2000, Shirai2003, Wheeler2011}. In addition, these ideas have been extended to nano-optical systems to describe photonic fields scattered by sub-wavelength nanostructures \cite{Yu2022, Gbur2001, you2020multiparticle, Aunon2014, Liu2022}. Even though the GSM originates from the classical theory of electromagnetic radiation, its versatility has enabled describing classical degrees of freedom of quantum optical systems \cite{nirala2023information, chen2021quantum, Hutter2020}. Specific examples include the modeling of the polarization, spectral, and orbital-angular-momentum properties of single and entangled photons \cite{nirala2023information, chen2021quantum, Hutter2020, You2023, Mandel1995}.

While the GSM fails to capture the intrinsic quantum properties of light, the quantum theory of electromagnetic radiation developed by Glauber and Sudarshan provides an elegant description for the excitation mode of the optical field \cite{glauber1963, sudarshan1963, Mandel1995}. Further, it provides the formalism to describe the quantum statistical properties of the light field and its quantum properties of coherence \cite{magana2019multiphoton, you2021observation, You2020a}. Indeed, these fundamental properties of photons are widely used to classify diverse kinds of light such as sunlight, laser radiation, and molecule fluorescence \cite{magana2019multiphoton, bhusal2021smart}. Notably, the most common type of light, classified under the umbrella of partially coherent light, can also be described using the classical theory of optical coherence \cite{Born1999, wolf2007introduction}. As such, partially coherent light beams are typically categorized as classical macroscopic optical systems \cite{Modi2012, Zurek1991a, frowis2018macroscopic}. Indeed, there has been interest in investigating the boundary between classical and quantum physics through the coherence properties of partially coherent optical beams \cite{de2012colloquium, Qian2015, Qian2017}. However, no previous studies have attempted to explore the constituent quantum multiphoton subsystems of classical partially coherent light.

Here, we establish a direct link between classical and quantum optics through the formulation of the quantum Gaussian-Schell (QGS) model. Remarkably, this model unveils the possibility of extracting the quantum multiphoton subsystems that constitute a classical partially coherent light field \cite{glauber1963, sudarshan1963, frowis2018macroscopic}. For the first time, we use photon-number-resolving detection to experimentally isolate the vacuum dynamics of a partially coherent light beam \cite{You2020a, 10.1063/5.0063294}. Furthermore, we discuss the quantum coherence properties of multiphoton wavepackets extracted from a classical light beam. We report experimental results for wavepackets with up to sixteen photons. The QGS model makes use of the complex-Gaussian statistical fluctuations inherent to partially coherent light \cite{Brown1956,MaganaLoaiza2016, Bromberg2010}. Indeed, our model explains how the quantum statistical fluctuations of the electromagnetic field give rise to the formation of macroscopic spatial correlations of partially coherent light. Surprisingly, we find that the quantum dynamics of the extracted multiphoton systems can be contrary to those of the classical system hosting them. This effect demonstrates the lack of a direct correspondence between the classical and quantum worlds \cite{GreggMarco}. Thus, we believe that the QGS model will have dramatic implications for quantum technologies \cite{willner2015optical, magana2019quantum,dell2006multiphoton,HongNP2023}.

\begin{figure}[!t]
    \centering
    \includegraphics[width=0.45\textwidth]{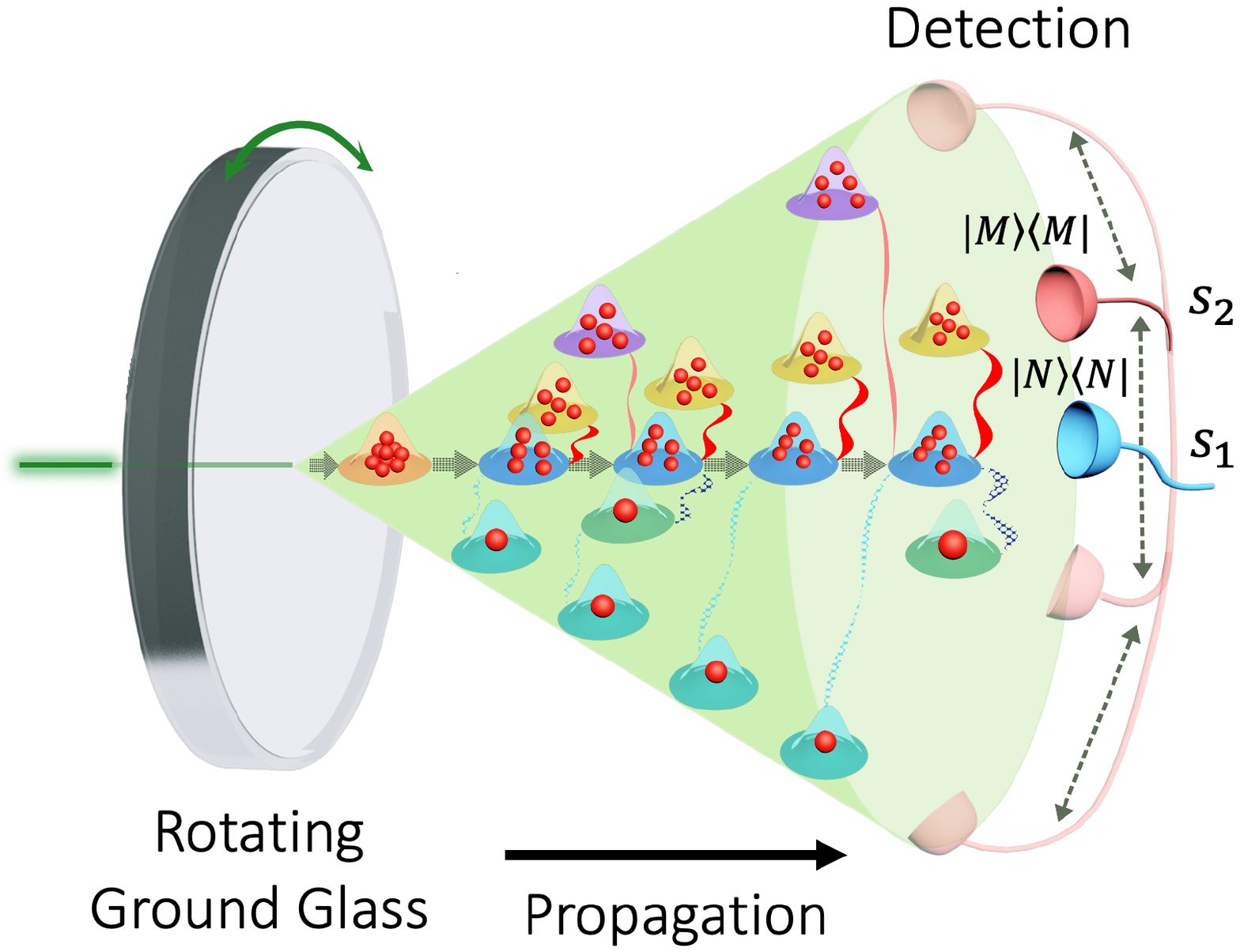}
    \caption{\textbf{Multiphoton wavepacket dynamics of partially coherent light.} This diagram illustrates 
    the generation of a partially coherent light field from a rotating ground glass \cite{Arecchi1965}. The macroscopic classical system is formed by many scattered multiphoton wavepakets exhibiting distinct quantum dynamics \cite{Smith2018}. For example, we illustrate bunching between five-photon wavepackets (red, solid) and anti-bunching between five- and one- photon wavepackets (blue, dotted). We isolate these multiphoton systems by performing projective measurements using photon-number-resolving detection \cite{You2020a, 10.1063/5.0063294}. In our experiment, we use two detectors, one of which is fixed at the center of the detection plane whereas the second is scanned through the wavefront of the scattered field. We conclude our experiment by performing multiphoton correlations between the two detectors. The outcomes of this experiment are described by the quantum Gaussian-Schell model.}
    \label{fig:figure1}
\end{figure}

As depicted in Fig. \ref{fig:figure1}, we are interested in describing quantum properties of coherence of the multiphoton systems that form partially coherent light. For the sake of generality, we consider a partially coherent light field produced by a rotating ground glass \cite{Smith2018, Arecchi1965}. This partially coherent field can be modeled through the indistinguishable superposition of coherent and thermal light beams. The coherent component results from the unaffected beam transmitted by the ground glass, whereas the thermal contribution is produced by many diffusers on the rotating ground glass that scatter the laser beam into many independent wavepackets \cite{Smith2018, Arecchi1965}. The collective properties of the resulting classical beam can be described by the following cross-spectral density function

\begin{equation}\label{eq1}
\begin{aligned}
    W(s_1,s_2) &=\braket{E^{(-)}(s_1)E^{(+)}(s_2)}\\
    &=\mu^*(s_1)\mu(s_2)+ \sqrt{\bar{n}(s_1)\bar{n}(s_2)}g(s_1-s_2)
\end{aligned} 
\end{equation}
for transverse-spatial positions $s_1$ and $s_2$. Here, $\mu(s_i)$ denotes the average electric-field amplitude $\braket{E^{(+)}(s_i)}$ at point $s_i$, $\bar{n}(s_i) \propto \text{Exp}\left[-|s_i|^2/\sigma_0\right]$ represents the mean photon number at point $s_i$, and $g(s_1-s_2) \propto \text{Exp}\left[-|s_1-s_2|^2/\sigma_1\right]$ is the normalized correlation term between the two points, where $\sigma_0$ and $\sigma_1$ are real, positive constants. The ensemble average $\braket{\cdot}$ is calculated with respect to the complex random variables $E^{(+)}(s_i)$, representing the electric field \cite{wolf2007introduction,korotkova2017random}. As discussed below, the inherent complex-Gaussian statistics of this beam enable the precise determination of its quantum statistics. To obtain the density matrix of the beam, we must first determine the probability density function for $E^{(+)}(s_1)$ and $E^{(+)}(s_2)$. This approach enables us to express the ensemble average $\braket{\cdot}$ in the form of an integral, incorporating the statistical characteristics of the electric field. The benefit of this method is the ability to compute expectation values for more complex structures than simple finite-degree polynomials of the electric field. For simplicity, we denote $E^{(+)}(s_1)\equiv \alpha$, $E^{(+)}(s_2)\equiv \beta$, $\mu(s_i) \equiv \mu_i$, $\bar{n}(s_i)\equiv\bar{n}_i$, and $g(s_1-s_2) \equiv g$.

Notably, we capture the quantum statistical properties of our field through the probability density of a real-Gaussian random $4$-vector $\boldsymbol{r}$ \cite{prince2012computer}. Despite the fact that this approach has not been used in quantum optics, it enables breaking the classical partially coherent field into its constituent multiphoton subsystems. Here, the corresponding probability density is 

\begin{equation}\label{eq2}
    P(\boldsymbol{r}) = \frac{1}{4\pi^2 \sqrt{|\boldsymbol{\Gamma}|}}e^{-\frac{1}{2}\left(\boldsymbol{r}-\boldsymbol{\mu}\right)^T\boldsymbol{\Gamma}^{-1}\left(\boldsymbol{r}-\boldsymbol{\mu}\right)},
\end{equation}
where $\boldsymbol{\mu} = \braket{\boldsymbol{r}}$ is the mean of $\boldsymbol{r}$, $\Gamma_{i,j} = \braket{r_i r_j} - \braket{r_i}\braket{r_j}$ is the covariance matrix of $\boldsymbol{r}$, and $|\cdot|$ represents the determinant. The critical aspect of Eq. (\ref{eq2}) is its exponent, where the correlations between random variables are manifested as products between them. Specifically, we calculate the mean vector as $\boldsymbol{\mu} = \left(\text{Re}[\mu_1], \text{Im}[\mu_1], \text{Re}[\mu_2], \text{Im}[\mu_2]\right)^\intercal$ and the covariance matrix as 
\begin{equation}\label{eq3}
    \boldsymbol{\Gamma} = \frac{1}{2}\begin{pmatrix}
        \bar{n}_1&0&g\sqrt{\bar{n}_1\bar{n}_2}&0\\
        0&\bar{n}_1&0&g\sqrt{\bar{n}_1\bar{n}_2}\\
        g\sqrt{\bar{n}_1\bar{n}_2}&0&\bar{n}_2&0\\
        0&g\sqrt{\bar{n}_1\bar{n}_2}&0&\bar{n}_2
    \end{pmatrix}. 
\end{equation}
Using Eq. (\ref{eq2}) and our expressions for $\boldsymbol{\mu}$ and $\boldsymbol{\Gamma}$, we can express the statistics of partially coherent light through its probability density function over coherent amplitudes $\alpha$ and $\beta$
\begin{widetext}
    \begin{equation}\label{eq4}
        P(\alpha,\beta) = \frac{1}{\pi^2\bar{n}_1\bar{n}_2(1-g^2)}\text{Exp}\left[-\frac{|\alpha - \mu_1|^2}{\bar{n}_1(1-g^2)} - \frac{|\beta - \mu_2|^2}{\bar{n}_2(1-g^2)} + 2g\frac{\text{Re}\left[(\alpha - \mu_1)^*(\beta - \mu_2)\right]}{\sqrt{\bar{n}_1\bar{n}_2}(1-g^2)}\right].
    \end{equation}
\end{widetext}
The amplitudes $\alpha$ and $\beta$ are centered around their respective means, with spatial coherence properties described by $g$. As a result, in the limit where $g\rightarrow 1$, the covariance matrix becomes degenerate, and the probability density function will instead become a probability distribution. Finally, when $s_1$ and $s_2$ are significantly different ($g\rightarrow 0$), the correlation term vanishes, resulting in two separate Gaussian distributions.

Remarkably, transitioning from the classical to the quantum description of our optical beam is now a straightforward process. This is because each instance of $E^{(+)}(s_i)$ in the classical ensemble corresponds to a coherent excitation \cite{ou2017quantum}. Thus, the density matrix of our beam, in terms of the excitation basis, can be written as
\begin{widetext}
    \begin{equation}
    \label{eq5}
        \hat{\rho}_{QGS}^D = \int d^2\alpha d^2\beta \frac{1}{\pi^2(\bar{n}_1+\bar{n}_1)^2(1-g^2)}\text{Exp}\left[-\frac{|\alpha - \tilde{\mu}_1|^2+|\beta - \tilde{\mu}_2|^2}{(\bar{n}_1 + \bar{n}_2)(1+g)}-g\frac{|\alpha - \beta + \tilde{\mu}_2-\tilde{\mu}_1|^2}{\left(1-g^2\right)(\bar{n}_1+\bar{n}_2)}\right] |\alpha c_\theta, \beta s_\theta\rangle\langle \alpha c_\theta, \beta s_\theta|,
    \end{equation}
\end{widetext}
where $c_\theta \equiv \cos(\theta) = \sqrt{\bar{n}_1/(\bar{n}_1+\bar{n}_2})$, $s_\theta \equiv \sin(\theta) = \sqrt{\bar{n}_2/(\bar{n}_1+\bar{n}_2})$, $\tilde{\mu}_1 = \mu_1 c_\theta^{-1}$, and $\tilde{\mu}_2 = \mu_2 s_\theta^{-1}$. Additionally, the superscript $D$ stands for ``detector perspective", which indicates that the information about the detectors' configuration is entirely contained in the density matrix $\hat{\rho}_{QGS}^D$ (see discussion in the SI). It is worth noticing that Eq. (\ref{eq5}) provides a suitable description to model the projection of a classical macroscopic light beam into its constituent multiphoton wavepackets. These projections are experimentally implemented using photon-number-resolving detection \cite{You2020a, 10.1063/5.0063294}. Further, Eq. (\ref{eq5}) enables describing the spatial correlation properties of isolated multiphoton subsystems. Interestingly, when ${\mu}_1={\mu}_2=0$, our beam reduces to the well-known GSM beam, which is characterized by photon statistics that always follow a thermal distribution \cite{wolf2007introduction, ou2017quantum}.

The above description of a QGS model source, while useful for understanding stationary instances of our beam, does not contain enough information to determine its evolution. This will require knowledge of our beam's underlying mode structure. Fortunately, we can utilize Eq. (\ref{eq5}), together with the spatial mode-structure of our light beam, to write its total quantum state as a functional integral (see SI for details)
\begin{equation}\label{eq6}
    \hat{\rho}_{QGS}^S = \int d\Sigma\text{ }|\alpha_0\rangle\langle\alpha_0|_\Sigma.
\end{equation}
Here, $S$ stands for ``state perspective", indicating that the information about the detectors' configuration is entirely contained in the detection operators, and that the density matrix $\hat{\rho}_{QGS}^S$ only contains information about the beam itself. We further note that $|\alpha_0\rangle_\Sigma$ is a coherent state characterized by a coherent amplitude $\alpha_0$ and a mode-structure given by $\hat{a} = \int ds \Sigma(s)\hat{a}(s)$. Additionally, the functional integral is over the random mode structure $\Sigma(s)$. Thus, we can use the following rule
\begin{equation}\label{eq7}
    \int d\Sigma\text{ } f\big(\boldsymbol{\Sigma}\big) = \int d\boldsymbol{\Sigma}\text{ } P\big(\boldsymbol{\Sigma}\big)f\big(\boldsymbol{\Sigma}\big),
\end{equation}
where $\boldsymbol{\Sigma}\equiv \big(\text{Re}[\Sigma(s_1)],\text{Im}[\Sigma(s_1)],...,\text{Re}[\Sigma(s_n)],\text{Im}[\Sigma(s_n)]\big)$ is a finite-dimensional vector of length $2n$, $f\big(\boldsymbol{\Sigma}\big)$ is a function of the $\Sigma(s_i)$, and $P\big(\boldsymbol{\Sigma}\big)$ is Eq. (\ref{eq2}) adapted to the variables present in $\boldsymbol{\Sigma}$. One can show that the statistical behavior of Eq. (\ref{eq6}) is identical to that of Eq. (\ref{eq5}), given appropriate choices of $\alpha_0$, ${\boldsymbol{\mu}}$ and $\boldsymbol{\Gamma}$. Moreover, Eq. (\ref{eq6}) has the additional benefit of a mode structure, which is necessary in determining the state's dynamical evolution through optical systems. Now, while the realization of the QGS model given by Eq. (\ref{eq6}) often proves to be more beneficial, Eq. (\ref{eq5}) remains valuable for its simplicity and as an important link between the classical GSM and the QGS model. Indeed, without the representation given in Eq. (\ref{eq5}), we would lack the basis on which to claim that Eq. (\ref{eq6}) represents the unique quantization of a partially coherent beam.

Now, we are in a position to explore how the underlying distribution of multiphoton wavepackets in the QGS model gives rise to the macroscopic properties of coherence in Eq. (\ref{eq1}). 
To do this, we compute the correlation properties between multiphoton wavepackets at the spatial locations $s_1$ and $s_2$. We can rewrite Eq. (\ref{eq5}) in the Fock state basis as $\hat{\rho}_{QGS}^D = \sum_{n,m,k,l=0}^\infty p_{\text{nmkl}}|n,m\rangle\langle k,l|$, which allows us to define a multiphoton wavepacket correlation function $\tilde{g}^{(2)}(N,M)$ as follows
\begin{equation}\label{eq8}
    \tilde{g}^{(2)}(N,M) = \frac{p_{\text{NMNM}}}{\left(\sum_{m=0}^\infty p_{\text{NmNm}}\right)\left(\sum_{n=0}^\infty p_{\text{nMnM}}\right)}.
\end{equation}
Here, $p_{\text{NMKL}} = \text{Tr}\left[\hat{\rho}_{QGS}^D\hat{n}_{\text{NMKL}}\right]$ is the probability associated with the Fock-projection operator $|N, M\rangle\langle K, L|$. Given that the probability of observing a specific multiphoton wavepacket is proportional to its number of occurrences, $\tilde{g}^{(2)}(N,M)$ effectively becomes the standard coherence function \cite{Mandel1995}. Specifically, $\tilde{g}^{(2)}(N,M)$ characterizes the coherence of $N$-photon wavepackets at the $s_1$-detector with $M$-photon wavepackets at the $s_2$-detector. Thus, the coherence function $\tilde{g}^{(2)}(N,M)$ is crucial for demonstrating the underlying nonclassical multiphoton coherence in partially coherent light sources, which can be critical for various applications in quantum information sciences \cite{dell2006multiphoton, magana2019quantum,bhusal2021smart,nirala2023information, you2020multiparticle,tame2013quantum, Yu2022, Gbur2001, Aunon2014, Liu2022}.

\begin{figure}[!htbp]
    \centering
    \includegraphics[width=0.47\textwidth]{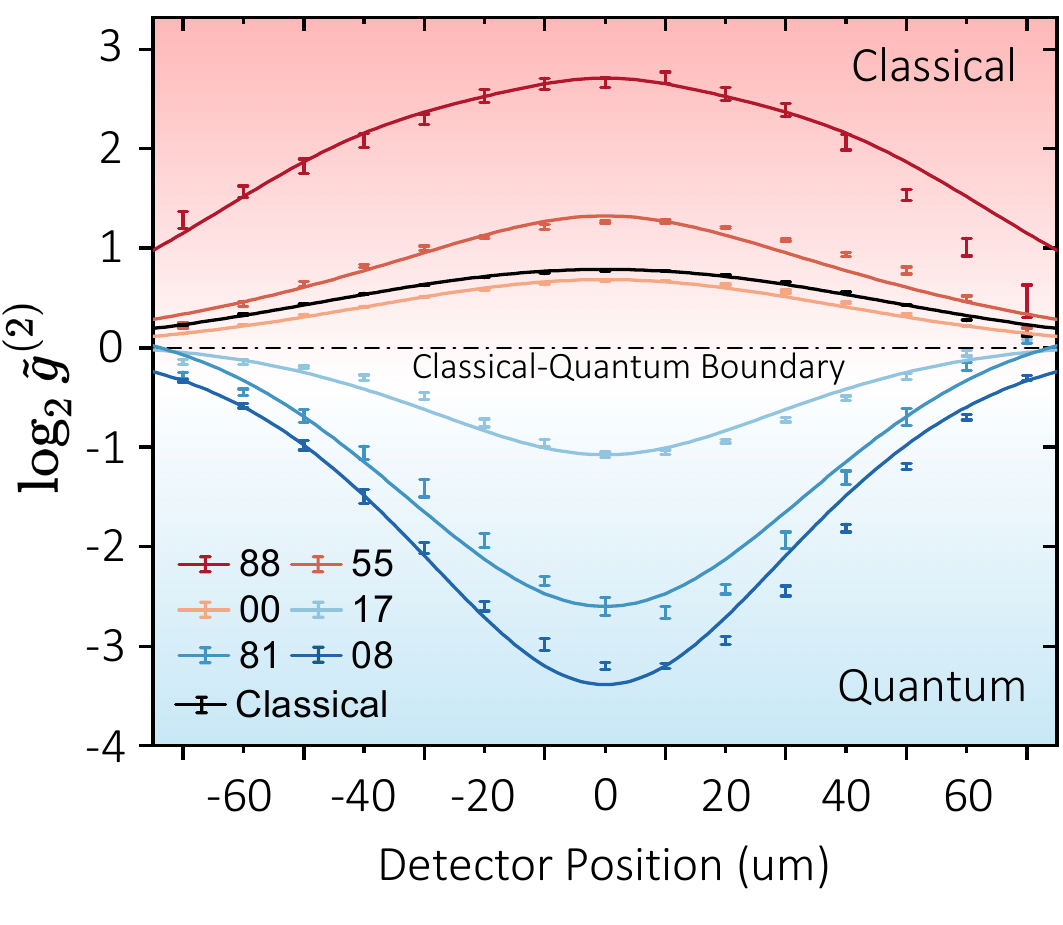}
    \caption{\textbf{Classical and Quantum Coherence of Multiphoton Wavepackets.} We present the experimental results of the process depicted in Figure \ref{fig:figure1}. Specifically, one detector was fixed at the center of the source, and the position of the second is given on the horizontal axis. On the vertical axis, we plot $\text{log}_2 \tilde{g}^{(2)}(N,M)$ for some of the constituent multiphoton wavepackets of the partially coherent light beam. We present these results for various choices of $N,M$ ranging from subsystems with zero to sixteen photons. Additionally, the classical $\text{log}_2 g^{(2)}$ for the partially coherent light beam is shown in black. The theoretical fitting of this data was accomplished using Eq. (\ref{eq8}), which remarkably describes the coherence properties of the extracted multiphoton subsystems. Interestingly, wavepackets whose $\tilde{g}^{(2)}$ have a positive logarithm exhibit classical properties of coherence, while those with a negative logarithm of $\tilde{g}^{(2)}$ show quantum coherence properties. Indeed, this kind of multiphoton subsystems cannot be described through the classical formulation of the Gaussian-Schell model \cite{wolf2007introduction, chen2021quantum, Peng2018, Yan2022, Liu2022a, Dutta2014, Rodenburg2014, Agrawal2000, Shirai2003, Wheeler2011, Yu2022, Gbur2001, you2020multiparticle, Aunon2014, Liu2022}.}
    \label{fig:figure2}
\end{figure}


We now experimentally verify the nonclassical properties of coherence of the constituent multiphoton subsystems of a partially coherent light field. Our technique for generating partially coherent light is depicted in Fig. \ref{fig:figure1}. Specifically, we generate a partially coherent light beam with a degree of second-order coherence $g^{(2)}(0)$ of $1.7$. This beam is passed through a beam splitter which produces two light fields that are measured by two photon-number-resolving (PNR) detectors. The first detector remains fixed at the center of the beam, whereas the second detector is moved along a transverse spatial axis. We report the correlation measurements for various choices of multiphoton wavepackets in Fig. \ref{fig:figure2}. While the trace of the classical partially coherent beam shows bunching properties, its constituent quantum multiphoton subsystems can exhibit very different dynamics. Surprisingly, the multiphoton subsystems with a vastly different number of particles exhibit antibunching, which is produced by the quantum nature of the light field \cite{Mandel1995}. However, the multiphoton subsystems with a similar number of particles show bunching effects like the classical hosting system. Interestingly, a similar behavior is observed for the correlation of the vacuum-fluctuation component of the field. Notably, the quantum Gaussian-Schell model introduced in this letter captures all the complex multiphoton dynamics hosted by partially coherent light. Furthermore, these results demonstrate that there is not a direct correpondence between the coherence properties of the classical partially coherent light field and its constituent multiphoton wavepackets \cite{GreggMarco}. In addition, these measurements show the possibility of extracting quantum multiphoton systems from classical light fields.

Our demonstration of the quantum Gaussian-Schell model establishes a direct relationship between the classical and quantum worlds \cite {Modi2012, Zurek1991a, frowis2018macroscopic, de2012colloquium, Qian2015, Qian2017}. This is achieved by extracting the constituent multiphoton quantum subsystems of a classical partially coherent light field. As a result of our predictions, we have for the first time experimentally isolated the vacuum and multiphoton dynamics of this kind of light field by implementing photon-number-resolving detection \cite{You2020a, 10.1063/5.0063294}. Our theory unveils surprising quantum coherence properties of multiphoton wavepackets, which have been observed in subsystems with up to sixteen photons. This uncovers the possibility of observing quantum properties within macroscopic classical objects. Furthermore, the quantum Gaussian-Schell model, leveraging complex-Gaussian statistical properties, elucidates the formation of macroscopic spatial correlations in partially coherent light \cite{Brown1956,MaganaLoaiza2016, Bromberg2010}. It also reveals surprising differences between the quantum dynamics of the constituent multiphoton wavepackets and their hosting classical system. Consequently, we believe that our work has profound implications for quantum imaging \cite{magana2019quantum,bhusal2021smart,nirala2023information}, quantum nanophotonics \cite{you2020multiparticle,tame2013quantum, Yu2022, Gbur2001, you2020multiparticle, Aunon2014, Liu2022}, and the preparation of multiparticle systems for quantum information science \cite{dell2006multiphoton, you2020multiparticle,magana2019quantum}.

Acknowledgments. - We acknowledge funding from the National Science Foundation through Grant No. CPS-2312086. We thank Dr. Ivan Agullo and Dr. Roberto de J. Le\'on-Montiel for very useful discussions.



\bibliography{main}

\clearpage

\onecolumngrid

\section{Derivation of the Quantum Gaussian-Schell Model (QGSM)}

The Gaussian-Schell Model (GSM) beam is a broad class of partially coherent beams \cite{wolf2007introduction, korotkova2017random}. A GSM beam can be described by a cross-spectral density (CSD) function \cite{schell1961multiple, CAI2017157}
\begin{equation}\label{eqS1}
    W(s_1,s_2) = \braket{E_{\text{th}}^{(-)}(s_1)E_{\text{th}}^{(+)}(s_2)} = \sqrt{\bar{n}(s_1)\bar{n}(s_2)}g(s_1-s_2),
\end{equation}
where $s_1,s_2$ are the transverse position variables, $\bar{n}(s_i) = \braket{E_{\text{th}}^{(-)}(s_i)E_{\text{th}}^{(+)}(s_i)}$, $g(s_1-s_2)$ is the degree of coherence between two spatial points, and $\braket{\cdot}$ is used to denote the ensemble average. Furthermore, the GSM beam requires that $\bar{n}(s_i) \propto \text{Exp}\left[-{|s_i|^2}/{\sigma_0}\right]$ and $g(s_1-s_2) \propto \text{Exp}\left[-{|s_1-s_2|^2}/{\sigma_1}\right]$ for some real, positive $\sigma_0,\sigma_1$. We note that the complex electric-field amplitude $E_{\text{th}}^{(+)}(s)$ obeys a standard complex normal random distribution as a result of the central limit theorem. Therefore, the beam will exhibit thermal statistics at each point in space, and as a result we obtain the relation $\braket{E_{\text{th}}^{(+)}} = 0$ \cite{ou2017quantum}.  

In this paper, however, we are interested in describing the more general case of a GSM beam overlapped with a coherent beam. We can describe this source by the complex electric-field $E^{(+)}(s) = E_{\text{th}}^{(+)}(s)+E^{(+)}_{\text{c}}(s)$, where $E_{\text{th}}^{(+)}(s)$ respresents the thermal electric-field amplitude of the GSM beam and $E^{(+)}_{\text{c}}(s)$ represents the coherent beam's electric-field amplitude. The coherent source obeys the relations $\braket{E^{(+)}_{\text{c}}(s)} = \mu(s)$ and $\braket{E^{(-)}_{\text{c}}(s_1)E^{(+)}_{\text{c}}(s_2)} = \mu^*(s_1)\mu(s_2)$, where $\mu(s)$ is the mean, or ``coherent amplitude" \cite{GlauberCoherentState}. We similarly require that $\mu(s_i) \propto \text{Exp}\left[-|s_i|^2/\sigma_0\right]$. Since the thermal and coherent fields both obey complex-Gaussian statistics, the resulting field $E^{(+)}(s)$ will similarly obey complex-Gaussian statistics. Assuming that the thermal and coherent fields are statistically independent of one-another, we can obtain the cross-spectral density of this new beam as
\begin{equation}\label{eqS2}
    \braket{E^{(-)}(s_1)E^{(+)}(s_2)} = \mu^*(s_1)\mu(s_2) + \sqrt{\bar{n}(s_1)\bar{n}(s_2)}g(s_1-s_2).
\end{equation}
Additionally, we have $\braket{E^{(+)}(s)} = \mu(s)$. 


In what follows, we will use the shorthands $E^{(+)}(s_1)\equiv \alpha$, $E^{(+)}(s_2)\equiv \beta$, $\mu(s_i) \equiv \mu_i$, $\bar{n}(s_i)\equiv\bar{n}_i$, and $g(s_1-s_2) \equiv g$ for simplicity. This allows us to rewrite Eq. (\ref{eqS1}) as $\braket{\alpha^*\beta} = \mu_1^*\mu_2 + g\sqrt{\bar{n}_1\bar{n}_2}$. Now, the probability density function for $\alpha,\beta$ is identical to the probability density function for $(\text{Re}[\alpha],\text{Im}[\alpha],\text{Re}[\beta],\text{Im}[\beta])\equiv (\alpha_1,\alpha_2,\beta_1,\beta_2)$, each of which is a real-Gaussian variable. From here, we can use the formula for the probability density of $2n$ real-Gaussian random variables represented by the vector $\boldsymbol{r}$, given by \cite{prince2012computer}
\begin{equation}\label{eqS3}
    P(\boldsymbol{r}) = \frac{1}{(2\pi)^n\sqrt{|\mathbf{\Gamma}|}}e^{-\frac{1}{2}(\boldsymbol{r}-\boldsymbol{\mu})^T \mathbf{\Gamma}^{-1}(\boldsymbol{r}-\boldsymbol{\mu})},
\end{equation}
where $\boldsymbol{\mu} = \braket{\boldsymbol{r}}$ is the mean vector and $\Gamma_{ij} = \braket{r_i r_j} - \braket{r_i}\braket{r_j}$ is the covariance matrix, and in our case $n=2$. So, if we can determine the values of $\boldsymbol{\mu}$ and $\boldsymbol{\Gamma}$, then we can determine the desired probability density function $P(\alpha,\beta)$. To accomplish this, we write $\alpha = \alpha_{\text{c}} + \alpha_{\text{th}}$ and $\beta = \beta_{\text{c}} + \beta_{\text{th}}$ where $\alpha_{\text{c}} = E^{(+)}_{\text{c}}(s_{1})$, $\beta_{\text{c}} = E^{(+)}_{\text{c}}(s_{2})$ and $\alpha_{\text{th}} = E^{(+)}_{\text{th}}(s_{1})$, $\beta_{\text{th}} = E^{(+)}_{\text{th}}(s_{2})$. 

We can see that $\boldsymbol{\mu} \equiv \left(\text{Re}[\mu_1], \text{Im}[\mu_1], \text{Re}[\mu_2], \text{Im}[\mu_2]\right)$, but determining $\boldsymbol{\Gamma}$ is less trivial. Since the covariance matrix elements of the coherent beam are all $0$, and since the thermal light and coherent light are independent of one-another, we see that the covariance matrix for $\alpha,\beta$ is identical to that of $\alpha_{\text{th}}\equiv \alpha_{\text{th},1} + i \alpha_{\text{th},2},\beta_{\text{th}}\equiv \beta_{\text{th},1} + i \beta_{\text{th},2}$. Now, recall that $\alpha_{\text{th},1},\beta_{\text{th},1}$ and $\alpha_{\text{th},2},\beta_{\text{th},2}$ are the in-phase and out-of-phase components respectively of $\alpha_{\text{th}}$, $\beta_{\text{th}}$. We can write $\alpha_{\text{th},1}\alpha_{\text{th},1} \equiv A(t)\cos^2(\omega t)$ and $\alpha_{\text{th},2}\alpha_{\text{th},2} \equiv A(t)\sin^2(\omega t)$ where $A(t)$ is the randomly-varying electric field amplitude and $\omega$ is the frequency of the light source. Noting that $A(t)$ fluctuates independently of the $\sin(\omega t)$ and $\cos(\omega t)$ terms, and that $\braket{\sin^2(\omega t)} = \braket{\cos^2(\omega t)}$, we can conclude that $\braket{\alpha_{\text{th},1}\alpha_{\text{th},1}} = \braket{\alpha_{\text{th},2}\alpha_{\text{th},2}} = \bar{n}_1/2$. Furthermore, we can write $\alpha_{\text{th},1}\alpha_{\text{th},2} \equiv A(t)\cos(\omega t)\sin(\omega t)$, and since $\braket{\cos(\omega t)\sin(\omega t)} = 0$, we get that $\braket{\alpha_{\text{th},1}\alpha_{\text{th},2}} = 0$. Similarly, we can show that $\braket{\beta_{\text{th},1}\beta_{\text{th},1}} = \braket{\beta_{\text{th},2}\beta_{\text{th},2}} = \bar{n}_2/2$ and $\braket{\beta_{\text{th},1}\beta_{\text{th},2}} = 0$. Identical arguments can be employed to show that $\braket{\alpha_1\beta_1} = \braket{\alpha_2\beta_2} = g\sqrt{\bar{n}_1\bar{n}_2}/2$ and $\braket{\alpha_1\beta_2} = \braket{\alpha_2\beta_1} = 0$. Putting all of this together, we conclude that 
\begin{equation}\label{eqS4}
    \boldsymbol{\Gamma} = \frac{1}{2}\begin{pmatrix}
        \bar{n}_1&0&g\sqrt{\bar{n}_1\bar{n}_2}&0\\
        0&\bar{n}_1&0&g\sqrt{\bar{n}_1\bar{n}_2}\\
        g\sqrt{\bar{n}_1\bar{n}_2}&0&\bar{n}_2&0\\
        0&g\sqrt{\bar{n}_1\bar{n}_2}&0&\bar{n}_2
    \end{pmatrix}.
\end{equation}
Then, by plugging $\boldsymbol{\mu},\mathbf{\Gamma}$ into Eq. (\ref{eqS3}), we get the probability density function
    \begin{equation}\label{eqS5}
        P(\alpha,\beta) = \frac{1}{\pi^2\bar{n}_1\bar{n}_2(1-g^2)}\text{Exp}\left[-\frac{|\alpha - \mu_1|^2}{\bar{n}_1(1-g^2)} - \frac{|\beta - \mu_2|^2}{\bar{n}_2(1-g^2)} + \frac{2\text{Re}\left[(\alpha - \mu_1)^*(\beta - \mu_2)\right]}{\sqrt{\bar{n}_1\bar{n}_2}(1-g^2)}\right].
    \end{equation}
Our beam is an ensemble average over many coherent beams where the ensemble distribution is described by Eq. (\ref{eqS5}). Therefore, the quantum state which represents our beam is given by 
\begin{equation}\label{eqS6}
    \hat{\rho}_{QGS}^D = \int d^2\alpha d^2\beta \text{ } P_{pc}(\alpha,\beta)|\alpha,\beta\rangle\langle\alpha,\beta|,
\end{equation}
where QGS stands for ``Quantum Gaussian-Schell," and D stands for ``Detector Perspective" for reasons which we will discuss in the next section. The expression for Eq. (\ref{eqS6}) can be simplified via a change-of-coordinates to yield
    \begin{equation}\label{eqS7}
        \hat{\rho}_{QGS}^D = \int d^2\alpha d^2\beta \frac{1}{\pi^2(\bar{n}_1+\bar{n}_1)^2(1-g^2)}\text{Exp}\left[-\frac{|\alpha - \tilde{\mu}_1|^2+|\beta - \tilde{\mu}_2|^2}{(\bar{n}_1 + \bar{n}_2)(1+g)}-g\frac{|\alpha - \beta + \tilde{\mu}_2-\tilde{\mu}_1|^2}{\left(1-g^2\right)(\bar{n}_1+\bar{n}_2)}\right] |\alpha c_\theta, \beta s_\theta\rangle\langle \alpha c_\theta, \beta s_\theta|,
    \end{equation}
where we have defined $c_\theta \equiv \cos(\theta) = \sqrt{\bar{n}_1/(\bar{n}_1+\bar{n}_2)}$, $s_\theta \equiv \sin(\theta) = \sqrt{\bar{n}_2/(\bar{n}_1+\bar{n}_2)}$, $\tilde{\mu}_1 = \mu_1 c_\theta^{-1}$, and $\tilde{\mu}_2 = \mu_2 s_\theta^{-1}$. Of course, if the two detectors are at the same spatial location then we would need to integrate over single-mode coherent states in Eq. (\ref{eqS6}) as opposed to two-mode coherent states. As a sanity-check, notice that in the limits where the detectors become close-together ($g\rightarrow 1$) or far-apart ($g\rightarrow 0$), we have
    \begin{equation}\label{eqS8}
        \begin{aligned}
            \lim_{g\rightarrow 0}\hat{\rho}_{QGS}^D &= \left[\int d^2\alpha \frac{1}{\pi \bar{n}_1}e^{-\frac{|\alpha-\mu_1|^2}{\bar{n}_1}}|\alpha\rangle\langle\alpha|\right]\otimes\left[\int d^2\beta \frac{1}{\pi \bar{n}_2}e^{-\frac{|\beta-\mu_2|^2}{\bar{n}_2}}|\beta\rangle\langle\beta|\right],\\
            \lim_{g\rightarrow 1}\hat{\rho}_{QGS}^D &= \int d^2\alpha \frac{1}{\pi(\bar{n}_1+\bar{n}_2)}e^{-\frac{\left|\alpha - \tilde{\mu_1}\right|^2}{\bar{n}_1+\bar{n}_2}}\left|\alpha c_\theta, \alpha s_\theta\right\rangle\left\langle\alpha c_\theta, \alpha s_\theta\right|.
        \end{aligned}
    \end{equation}
In the $g\rightarrow 0$ limit, the state approaches the tensor product of two independent partially coherent beams. In the $g\rightarrow 1$ limit, however, one of the integration variables is eliminated and $\hat{\rho}_{QGS}^D$ mimics a partially coherent beam which has been split into two modes. Now, we can compute the density matrix elements of $\hat{\rho}_{QGS}^D$ by taking the expectation value of the operator $\hat{n}_{\text{NMKL}} = |N,M\rangle\langle K,L|$. Doing so yields
\begin{equation}\label{eqS9}
    \begin{aligned}
\text{Tr}\left[\hat{\rho}_{QGS}^D\hat{n}_{\text{NMKL}}\right] &= \int d^2\alpha d^2\beta P(\alpha_1,\alpha_2,\beta_1,\beta_2)\text{Exp}\left[-|\alpha|^2-|\beta|^2\right]\frac{\alpha^N\alpha^{*K}\beta^M\beta^{*L}}{\sqrt{N!K!M!L!}}\\
        &=\frac{1}{\sqrt{N!K!M!L!}}\int d^2\alpha d^2\beta P(\alpha_1,\alpha_2,\beta_1,\beta_2)\text{Exp}\left[-\alpha_1^2-\alpha_2^2-\beta_1^2-\beta_2^2\right]\\
        &\text{ }\text{ }\text{ }\text{ }\text{ }\text{ }\text{ }\text{ }\text{ }\text{ }\text{ }\text{ }\text{ }\text{ }\text{ }\text{ }\text{ }\text{ }\text{ }\text{ }\text{ }\text{ }\text{ }\text{ }\text{ }\text{ }\text{ }\text{ }\text{ }\text{ }\text{ }\text{ }\text{ }\text{ }\text{ }\times(\alpha_1+i\alpha_2)^N(\alpha_1-i\alpha_2)^K(\beta_1+i\beta_2)^M(\beta_1-i\beta_2)^L\\
        &= \frac{1}{\sqrt{N!K!M!L!}}\sum_{n=0}^N\sum_{k=0}^K\sum_{m=0}^M\sum_{l=0}^L\binom{N}{n}\binom{K}{k}\binom{M}{m}\binom{L}{l} i^{n-N+m-M-k+K-l+L} \\
        &\text{ }\text{ }\text{ }\text{ }\times \int d\alpha_1d\alpha_2d\beta_1d\beta_2 P(\alpha_1,\alpha_2,\beta_1,\beta_2)e^{-\alpha_1^2-\alpha_2^2-\beta_1^2-\beta_2^2}\alpha_1^{n+k}\alpha_2^{N+K-n-k}\beta_1^{m+l}\beta_2^{M+L-m-l}\\
        &= \frac{1}{\sqrt{N!K!M!L!}}\sum_{n=0}^N\sum_{k=0}^K\sum_{m=0}^M\sum_{l=0}^L\binom{N}{n}\binom{K}{k}\binom{M}{m}\binom{L}{l} i^{n-N+m-M-k+K-l+L} \\
        &\text{ }\text{ }\text{ }\text{ }\times \int d\alpha_1 d\beta_1 \tilde{P}(\alpha_1,\beta_1)e^{-\alpha_1-\beta_1}\alpha_1^{n+k}\beta_1^{m+l}\int d\alpha_2 d\beta_2 \tilde{P}(\alpha_2,\beta_2)e^{-\alpha_2-\beta_2}\alpha_2^{N+K-n-k}\beta_2^{M+L-m-l},
        \end{aligned}
\end{equation}
where in the last line we have defined
\begin{equation}\label{eqS10}
    \tilde{P}(\alpha,\beta) = \frac{1}{\pi\sqrt{\bar{n}_1\bar{n}_2(1-g^2)}}\text{Exp}\left[-\frac{(\alpha - \mu)^2}{\bar{n}_1(1-g^2)} - \frac{(\beta - \eta)^2}{\bar{n}_2(1-g^2)}+\frac{2g(\alpha - \mu)(\beta - \eta)}{\sqrt{\bar{n}_1\bar{n}_2}(1-g^2)}\right].
\end{equation}
Here, $\mu = \text{Re}[\mu_{1}],\text{Im}[\mu_{1}]$ and $\eta = \text{Re}[\mu_{2}],\text{Im}[\mu_{2}]$ depending on which integral is under consideration. In order to evaluate Eq. (\ref{eqS9}), we need to evaluate the more general integral 
\begin{equation}\label{eqS11}
    \int d\alpha d\beta \alpha^N \beta^M \text{Exp}\left[-x \alpha^2 - y \beta^2 + 2z \cdot \alpha\beta + u\alpha + v\beta + w\right].
\end{equation}
where we have defined $x = \frac{1+\bar{n}_1(1-g^2)}{\bar{n}_1(1-g^2)}$, $y = \frac{1+\bar{n}_2(1-g^2)}{\bar{n}_2(1-g^2)}$, $z = \frac{g}{\sqrt{\bar{n}_1\bar{n}_2}(1-g^2)}$, $u = 2(\mu (x-1) - \eta z)$, $v = 2(\eta (y-1) - \mu z)$, and $w = 2\mu\eta z - (x-1)\mu^2 - (y-1)\eta^2$ for ease of notation. Then we perform the substitution $\gamma = \alpha - \frac{y}{z}\beta$, resulting in the integral
\begin{equation}\label{eqS12}
    \begin{aligned}
        \left(\frac{z}{y}\right)^{M+1} \sum_{k=0}^M\binom{M}{k}(-1)^{k}\left[\int d\alpha\text{ }\alpha^{N+M-k}e^{-\left(x-\frac{z^2}{y}\right)\alpha^2 - \left(\frac{z v}{y}-u\right)\alpha}\right]\left[\int d\beta\text{ }\beta^k e^{-\left(\frac{z^2}{y}\right)\gamma^2 - \left(\frac{z v}{y}\right)\gamma}\right]e^w.
    \end{aligned}
\end{equation}
Evaluating this analytically will mean utilizing the even more general integral formula
\begin{equation}\label{eqS13}
\begin{aligned}
    f(a,b,n) &= \int dq\text{ } q^n e^{-a q^2 - b q}\\
    &= \frac{1}{2 a^{\frac{n}{2}+1}}\left[(-1+(-1)^n)b\Gamma\left(1+\frac{n}{2}\right) \text{}_1F_1\left(1+\frac{n}{2},\frac{3}{2},\frac{b^2}{4a}\right) + (1+(-1)^n)\sqrt{a}\Gamma\left(\frac{1+n}{2}\right)\text{}_1F_1\left(\frac{1+n}{2},\frac{1}{2},\frac{b^2}{4a}\right)\right].
\end{aligned}
\end{equation}
With this in hand, we can write the preceeding integral as
\begin{equation}\label{eqS14}
    \left(\frac{z}{y}\right)^{M+1} \sum_{k=0}^M\binom{M}{k}(-1)^{k}f\left(x-\frac{z^2}{y},\frac{z v}{y} - u, N+M-k\right)f\left(\frac{z^2}{y},\frac{z v}{y},k\right)e^w.
\end{equation}
Then, denoting this expression as $p(N,M,\bar{n}_1,\bar{n}_2,g,\mu,\eta)\left(\pi\sqrt{\bar{n}_1\bar{n}_2(1-g^2)}\right)$, we ultimately get the detailed expression for Eq. (\ref{eqS9}) given by
\begin{equation}\label{eqS15}
    \begin{aligned}
        \text{Tr}\left[\hat{\rho}_{QGS}^D\hat{n}_{\text{NMKL}}\right] &=\frac{1}{\sqrt{N!M!K!L!}}\sum_{n=0}^N\sum_{m=0}^M\sum_{k=0}^K\sum_{l=0}^L\binom{N}{n}\binom{M}{m}\binom{K}{k}\binom{L}{l} i^{n-N+m-M-k+K-l+L} \\
        &\text{ }\text{ }\text{ }\text{ }\times p(n+k,m+l,\bar{n}_1,\bar{n}_2,g,\text{Re}[\mu_1],\text{Re}[\mu_2])\\
        &\text{ }\text{ }\text{ }\text{ }\times p(N+K-n-k,M+L-m-l,\bar{n}_1,\bar{n}_2,g,\text{Im}[\mu_1],\text{Im}[\mu_2])
    \end{aligned}
\end{equation}
for the density matrix elements of $\hat{\rho}_{QGS}^D$. While a simpler expression would be desirable, we note that Eq. (\ref{eqS15}) can be numerically evaluated without difficulty.

Here, we note that the QGS model given by Eq. (\ref{eqS6}) is limited to the study of a stationary source. In order to study the free-space dynamics of a QGS model source, we must describe our beam with a mode structure. Fortunately, we can use Eq. (\ref{eqS6}) together with the knowledge of how QGS model beams are generated to determine that mode structure. As discussed above, a QGS beam can be generated by illuminating a randomly-changing surface with a coherent beam of light. The resulting state can be represented by a functional integral
\begin{equation}\label{eqS16}
    \hat{\rho}_{QGS}^S = \int d\Sigma\text{ }|\alpha_0\rangle\langle\alpha_0|_\Sigma,
\end{equation}
where S stands for ``state perspective", and $|\alpha_0\rangle_\Sigma$ is a coherent state with coherent amplitude $\alpha_0$ and a mode-structure given by 
\begin{equation}\label{eqS17}
    \hat{a} = \int ds \text{ } \Sigma(s)\hat{a}(s).
\end{equation}
Here, $\Sigma(s)$ is the random intensity profile. In our case, the functional integral is given by
\begin{equation}\label{eqS18}
    \int d\Sigma\text{ } f(\Sigma_1,...,\Sigma_n) = \int d^2\Sigma_1...d^2\Sigma_n\text{ } \frac{1}{(2\pi)^n\sqrt{|\mathbf{\Gamma}|}}e^{-\frac{1}{2}(\boldsymbol{r}-\boldsymbol{\mu})^T \Gamma^{-1}(\boldsymbol{r}-\boldsymbol{\mu})}f(\Sigma_1,...,\Sigma_n),
\end{equation}
where $\Sigma_i \equiv \Sigma(s_i)$, $f(\Sigma_1,...,\Sigma_n)$ is a function of many $\Sigma_i$, $\boldsymbol{r} \equiv \Big(\text{Re}[\Sigma_1],\text{Im}[\Sigma_1],...,\text{Re}[\Sigma_n],\text{Im}[\Sigma_n]\Big)$, $\boldsymbol{\mu} = \braket{\boldsymbol{r}}$, $\mathbf{\Gamma}$ is $\boldsymbol{r}$'s covariance matrix, and $|\cdot|$ is the determinant operation. Additionally, we note that the integration operator $\int d\Sigma$ commutes with integrals over finitely many position variables of the form $\int ds_1...ds_n$. The statistical behavior of this state is then identical to that of Eq. (\ref{eqS6}) for suitable choice of $\alpha_0$, ${\boldsymbol{\mu}}$ and $\boldsymbol{\Gamma}$, but it has the additional benefit of a mode structure, which can be used to determine the dynamical evolution of the state through an optical system. Therefore, the realization of the QGS model given by Eq. (\ref{eqS16}) will, in many cases, be more useful than that of Eq. (\ref{eqS6}). However, the latter still finds its utility as a simplified description, as well as an important link between the classical GSM and the QGS model presented in Eq. (\ref{eqS16}). Indeed, without the representation given in Eq. (\ref{eqS6}), we would not have the grounds on which to claim that Eq. (\ref{eqS16}) represents the unique quantization of a GSM beam.

\section{Correlation Dynamics of Multiphoton Wavepackets}

All light sources consist of photons, which can be described as a statistical superposition of the various possible excitations of the source's mode structure. Each of these excitations, containing a fixed number of photons, will henceforth be referred to as a multiphoton wavepacket. The quantum state representing a multiphoton wavepacket containing exactly $N$ photons is the Fock state $|N\rangle$. This state is characterized by creation and annihilation operators that are defined in accordance with the source's mode structure. Consequently, the operator responsible for measuring the probability of detecting an $N$-photon wavepacket is given by $\hat{n}_{\text{N}}\equiv |N\rangle\langle N|$. Given that the probability of any event is intrinsically linked to how frequently it occurs, a practical approach to experimentally measuring this probability involves photon-number-resolving (PNR) detection. By post-selecting on the photon-number-resolved detection, we can obtain a more accurate estimation of the event's probability.

With our understanding of light's underlying structure, it naturally leads to questions about the behavior of multiphoton wavepackets in specific geometric circumstances. For instance, what would be the outcome on the multiphoton wavepacket distribution if we were to measure it using a point-detector, as opposed to a detector tailored to the mode-structure of the source? There are two viable approaches to address this question. The first involves leveraging existing knowledge about the overall state to define a quantum framework for the point detector and subsequently determine the distribution of these quanta. Alternatively, and perhaps more feasibly in the absence of such information, is to identify a point-detector equivalent of the operator $\hat{n}_{\text{N}}$. In this latter approach, one could utilize this analogue to infer the quanta distribution relevant to the first method. Given its broader applicability, it's worthwhile to delve deeper into this discussion.

Let us begin with the mode operators of the total state, $\hat{a}$, such that $|N\rangle = \frac{\left(\hat{a}^\dagger\right)^N}{\sqrt{N!}}|0\rangle$. One can show that
\begin{equation}\label{eqS19}
    \left(\hat{a}^\dagger\right)^n \left(\hat{a}\right)^n = \sum_{k=n}^\infty \frac{k!}{(k-n)!}|k\rangle\langle k|.
\end{equation}
Then, we see that
\begin{equation}\label{eqS20}
\begin{aligned}
    :\left[e^{-\hat{n}}\right]:\text{ } = \text{ }:\left[e^{-\hat{a}^\dagger \hat{a}}\right]:\text{ }&=  \sum_{n=0}^\infty \frac{(-1)^n}{n!}\left(\hat{a}^\dagger\right)^n \left(\hat{a}\right)^n= \sum_{n=0}^\infty \frac{(-1)^n}{n!}\sum_{k=n}^\infty \frac{k!}{(k-n)!}|k\rangle\langle k|\\
    &= \sum_{n=0}^\infty\sum_{k=n}^\infty (-1)^n \binom{k}{n}|k\rangle\langle k|= \sum_{n=0}^\infty\sum_{k=0}^\infty (-1)^n\binom{n+k}{k}|n+k\rangle\langle n+k|\\
    &= \sum_{n=0}^\infty \left(\sum_{k=0}^n (-1)^k\binom{n}{k}\right)|n\rangle\langle n|= |0\rangle\langle 0|,
\end{aligned}   
\end{equation}
where $\hat{n} = \hat{a}^\dagger\hat{a}$ is the number operator and $:\cdot:$ is the normal-ordering prescription. In the last line, we have used the identity $0^n = (1 - 1)^n = \sum_{k=0}^n (-1)^k\binom{n}{k}$. As a result, we also conclude that $\hat{n}_{\text{NK}} \equiv |N \rangle\langle K|$ can be written as
\begin{equation}\label{eqS21}
    \hat{n}_{\text{NK}} = \frac{1}{\sqrt{N!K!}}:\left[\left(\hat{a}^\dagger\right)^N \left(\hat{a}\right)^K e^{-\hat{a}^\dagger \hat{a}}\right]:.
\end{equation}
Therefore, just as the point-detector analogue at position $s$ of $\hat{a}$ is $\hat{a}(s)$, the analogue of $\hat{n}_{\text{NK}}$ is given by
\begin{equation}\label{eqS22}
    \hat{n}_{\text{NK}}(s) = \frac{1}{\sqrt{N!K!}}:\left[\left(\hat{a}^\dagger(s)\right)^N \left(\hat{a}(s)\right)^K e^{-\hat{a}^\dagger (s)\hat{a}(s)}\right]:.
\end{equation}
It is now worth recalling our definitions from the previous section, where we defined $\hat{\rho}_{QGS}^D$ and $\hat{\rho}_{QGS}^S$. The first representation $\hat{\rho}_{QGS}^D$, the so-called detector perspective, is what one would obtain by defining a notion of quanta for two point detectors and then interpreting their statistical distribution from known behavior. The second representation $\hat{\rho}_{QGS}^S$, the so-called state perspective, will yield the same distribution using the Fock-projection operators, which we have just derived. Specifically, we can write
\begin{equation}\label{eqS23}
    \text{Tr}\left[\hat{\rho}_{QGS}^D\hat{n}_{\text{NMKL}}\right] = \text{Tr}\left[\hat{\rho}_{QGS}^S:\hat{n}_{\text{NK}}(s_1)\hat{n}_{\text{ML}}(s_2):\right].
\end{equation}
The use of the terms `detector perspective' and `state perspective' now should become clear. When altering the detector configuration from the detector's perspective, the change is reflected in the state $\hat{\rho}_{QGS}^D$ in Eq. (\ref{eqS23}). However, from the state's perspective, it is the detector's measurement operator $:\hat{n}_{\text{NK}}(s_1)\hat{n}_{\text{ML}}(s_2):$, that changes, not the state $\hat{\rho}_{QGS}^S$. Therefore, in the detector's view, the measurement operator remains constant while the state changes. Conversely, from the state's standpoint, it is the detector that is changing. Generally, the state perspective offers a more versatile and intuitive understanding of the total system's geometry. However, the detector perspective provides a clearer insight into the multiphoton wavepacket distribution of the measurement. While the detector perspective state can be deduced using the state perspective in conjunction with Eq. (\ref{eqS23}), the reverse is often not true.  Eq. (\ref{eqS6}) can be derived from Eq. (\ref{eqS16}), but not vice-versa, suggesting that Eq. (\ref{eqS16}) is the more comprehensive QGS model between the two states presented in the last section. We note that the distinction between the detector perspective and the state perspective is not equivalent to the difference between the Schr\"odinger and Heisenberg pictures of time evolution. Indeed, time evolution can only take place in the state perspective and no dynamics can occur in the detector perspective. 

Now, with this understanding multiphoton wavepacket measurements using point-detection, we turn our attention to an intriguing aspect of such wavepackets originating from a QGS model source, particularly their spatial intensity correlations. We start by defining the multiphoton wavepacket correlation function as
\begin{equation}\label{eqS24}
    \tilde{g}^{(2)}(N,M) = \frac{p_{\text{NMNM}}}{\left(\sum_{m=0}^\infty p_{\text{NmNm}}\right)\left(\sum_{n=0}^\infty p_{\text{nMnM}}\right)},
\end{equation}
where $p_{\text{NMKL}} = \text{Tr}\left[\hat{\rho}_{QGS}^D\hat{n}_{\text{NMKL}}\right]$. Given that the probability of observing a specific multiphoton wavepacket is proportional to its number of occurrences, we see that $\tilde{g}^{(2)}(N,M)$ is the standard correlation function between occurences of $N$-photon wavepackets at the $s_1$-detector and $M$-photon wavepackets at the $s_2$-detector. These correlations can be calculated using Eq. (\ref{eqS15}), revealing very interesting results.

Specifically, we observe a strong correlation ($\tilde{g}^{(2)}(N,M) > 1$) when $N = M$ and $s_1\approx s_2$. Conversely, a significant anticorrelation ($\tilde{g}^{(2)}(N,M) < 1$) emerges when $N$ greatly differs from $M$.  When $s_1$ and $s_2$ are distant from each other, these correlations and anticorrelations monotonically approach $1$ in a nearly Gaussian manner. These theoretical predictions are in strong agreement with our experiment observations. The observed anticorrelations, in particular, highlight the non-classical behaviors originating from an otherwise classically-behaving source of light.

\end{document}